\documentclass[conference]{IEEEtran}
\IEEEoverridecommandlockouts
\usepackage{cite}
\usepackage{amsmath,amssymb,amsfonts}
\usepackage{algorithmic}
\usepackage{graphicx}
\usepackage{textcomp}
\usepackage{xcolor}
\usepackage{booktabs}
\usepackage{listings}
\usepackage{multirow}
\usepackage{float}
\usepackage{dblfloatfix}
\usepackage{xspace}

\def\BibTeX{{\rm B\kern-.05em{\sc i\kern-.025em b}\kern-.08em
    T\kern-.1667em\lower.7ex\hbox{E}\kern-.125emX}}
\begin{document}
\onecolumn
\begin{flushleft}
“© 2019 IEEE.  Personal use of this material is permitted. Permission from IEEE must be obtained for all other uses, in any current or future media, including reprinting/republishing this material for advertising or promotional purposes, creating new collective works, for resale or redistribution to servers or lists, or reuse of any copyrighted component of this work in other works.”
\end{flushleft}.
\twocolumn

\title{Efficient Fault Injection based on Dynamic HDL Slicing Technique
\thanks{This research was supported by project RESCUE funded from the European Union’s Horizon 2020 research and innovation programme under the Marie Sklodowaska-Curie grant agreement No 722325.}
}

%

\author{\IEEEauthorblockN{Ahmet Cagri Bagbaba\IEEEauthorrefmark{1}\IEEEauthorrefmark{2},
Maksim Jenihhin\IEEEauthorrefmark{2}, Jaan Raik\IEEEauthorrefmark{2}, Christian Sauer\IEEEauthorrefmark{1}}
\IEEEauthorblockA{\IEEEauthorrefmark{1}Cadence Design Systems,
Munich, Germany; \IEEEauthorrefmark{2} Tallinn University of Technology, Tallinn, Estonia \\
Email: \IEEEauthorrefmark{1}\{abagbaba, sauerc\}@cadence.com,
\IEEEauthorrefmark{2}\{maksim.jenihhin, jaan.raik\}@taltech.ee}}

\maketitle

\begin{abstract}
This work proposes a fault injection methodology where Hardware Description Language (HDL) code slicing is exploited to prune fault injection locations, thus enabling more efficient campaigns for safety mechanisms evaluation. In particular, the dynamic HDL slicing technique provides for a highly collapsed critical fault list and allows avoiding injections at redundant locations or time-steps. Experimental results show that the proposed methodology integrated into commercial tool flow doubles the simulation speed when comparing to the state-of-the-art industrial-grade EDA tool flows.
\end{abstract}


\section{Introduction}

During the design of ISO26262~\cite{ISO} compliant chips, designers need to evaluate effectiveness of the design to deal with random hardware faults. This is usually done by the fault injection simulations. The goal of a fault injection experiment is to exercise the system's fault protection capabilities. Faults which cause the system to fail in the absence of fault detection capabilities are defined to be \emph{critical}. A \emph{critical fault}, if undetected in presence of fault processing mechanism, will result in a failure of the system under test. Using critical faults to estimate fault coverage eliminates the possibility of fault injection experiments to produce no errors. Several approaches to generate the critical fault list to speed up the fault injection campaigns have been proposed. However, to the best of the authors' knowledge this is the first work where dynamic HDL slicing has been implemented in order to minimize the number of fault injections. The main contributions proposed by this work as follows:

\begin{itemize}
  \item Dynamic slicing on HDL for critical fault list generation.
  \item Language-agnostic RTL fault injection.
\end{itemize}

As a result, significant speed-up of the fault injection simulation is achieved. Experimental results show that the proposed methodology doubles the simulation speed when comparing to the state-of-the-art optimizations based on static cone approach. Only fault model implemented in this paper is based on single-clock-cycle bit-flip faults within the RTL registers. This fault model is targeting single Single-Event-Upsets (SEUs) in all the flip-flops of the design. The proposed methodology is demonstrated on Cadence tools but it remains applicable to other tool flows as well.

In the majority of the published literature~\cite{7684076},~\cite{8347235} fault location and fault insertion time are randomly selected as opposed to the methodology explained in this paper. In addition, previous works~\cite{newJ} have demonstrated that with randomly selected fault lists the ratio of faults which do not produce errors may range as low as 2 to 8 per cent, depending on the system under simulation. Therefore, minimization of fault injection locations has a potential to reduce the time of the fault injection campaign significantly while allowing injection and simulation of a considerably larger number of relevant faults. Additionally, dynamic slicing technique is used in~\cite{6549113},~\cite{6233020} for statistical bug localization in RTL. Different from the works listed above, this paper proposes a dynamic HDL slicing based technique that implicitly covers the golden run fault collapsing, thereby significantly speeding up the fault injection process.

\begin{figure}[b]
\centerline{\includegraphics[scale=0.62]{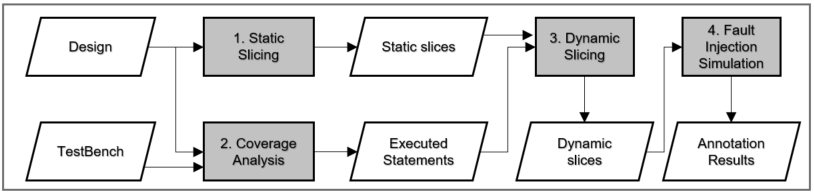}}
\caption{Proposed Dynamic HDL slicing based fault injection methodology.}
\label{flow}
\end{figure}

\section{Proposed Methodology}
\begin{figure}[t]
\centerline{\includegraphics[scale=0.68]{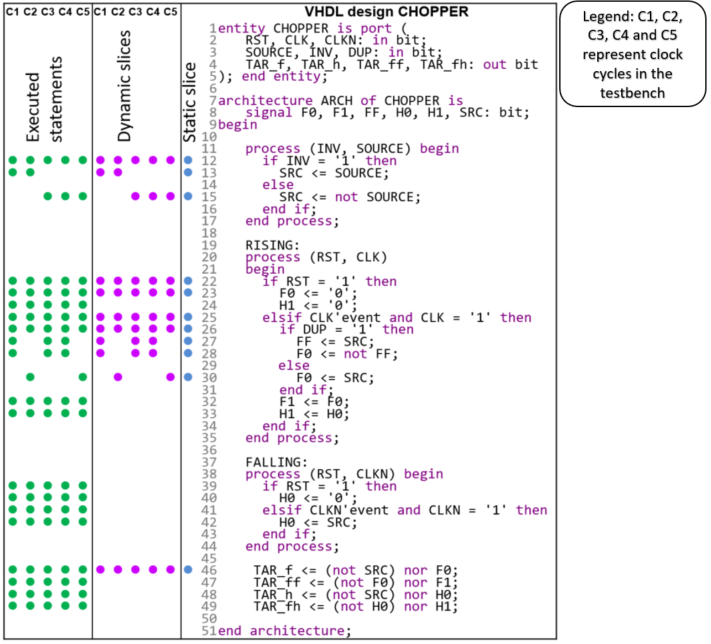}}
\caption{HDL slicing on a motivational example chopper~\cite{Clarke2002ProgramSF}.}
\label{mot}
\end{figure}

\begin{table*}[b]
\small
\centering
\caption{Fault Injection Campaign Results for Chopper and Simple\_SPI Designs}
\label{bigtable}
\resizebox{.72\textwidth}{!}{%
\begin{tabular}{|c|c|c|c|c|}
\hline
Design Name & \multicolumn{2}{c|}{\textbf{chopper}} & \multicolumn{2}{c|}{\textbf{simple\_spi}} \\ \hline
Optimization type & \textbf{\begin{tabular}[c]{@{}c@{}}Static Slice\end{tabular}} & \textbf{\begin{tabular}[c]{@{}c@{}}Dynamic Slice\end{tabular}} & \textbf{\begin{tabular}[c]{@{}c@{}}Static Slice\end{tabular}} & \textbf{\begin{tabular}[c]{@{}c@{}}Dynamic Slice\end{tabular}} \\ \hline
Observation list & \multicolumn{2}{c|}{tar\_f} & \multicolumn{2}{c|}{dat\_o} \\ \hline
Fault target & F0, FF & dynamic slices & mem{[}{]}{[}{]} & dynamic slices \\ \hline
Total number of injected faults & 410 & 255 & 210080 & 960 \\ \hline
Number of detected faults & 220 & 137 & 1696 & 609 \\ \hline
Number of undetected faults & 190 & 118 & 208384 & 351 \\ \hline
Total CPU time of overall regression & 1.33s & 1.2s & 171.5s & 15.2s \\ \hline
\end{tabular}
}
\end{table*}
The proposed methodology is outlined in Fig.~\ref{flow}. We explain the details of the methodology in the following paragraphs by using a motivational example depicted in Fig.~\ref{mot}.

Static slice(1) shows the dependency between HDL statements~\cite{Iwaihara96programslicing}. Static slice column in Fig.~\ref{mot} shows the HDL statements which are in static slice of \emph{TAR\_F} output. Fig.~\ref{mot} also implies that, static slice does not depend on clock cycles (shown as C1, C2, C3, C4 and C5). In this work, Cadence{\textregistered} JasperGold Formal Verification Platform is used to calculate backward static slice.

In parallel to static slicing step, the RTL design is simulated in Cadence{\textregistered} Xcelium{\texttrademark} simulator to dump and analyse coverage data(2). In this step, we dump coverage data for each clock cycle so that we can find what statements in the RTL are executed for each clock cycle. In the proposed methodology, one clock cycle defines the size of our dynamic slice. We use code coverage which measures how thoroughly a testbench exercises the lines of HDL code. At the end of this step, we generate executed statements to use it in the next step. Fig.~\ref{mot} shows executed statements for five clock cycles (C1, C2, C3, C4, C5).

Dynamic slicing(3), as it is implemented here, includes those statements that actually affect the value of a variable for a particular set of inputs of the RTL so it is computed on a given input~\cite{Korel:1988:DPS:56378.56386}. It provides more narrow slices than static slice and consists of only the statements that effect the value of a variable for a given input. In a nutshell, dynamic slice is the intersection of static slice and executed statements as in the Fig.~\ref{mot}. For instance, during the time window C5, register \emph{FF} (Line 27) is not in dynamic slice meaning that we do not need to inject fault in \emph{FF} at C5 time window. Dynamic slice gives us critical faults and eliminates those faults that are not critical. In this way, we manage to reduce fault list by injecting only critical faults. This provides significant speed-up in the fault injection simulation time as each injected fault increases total run time of fault injection campaign.

For the fault injection simulation step(4), we use Cadence{\textregistered} Xcelium{\texttrademark} Fault Simulator. Fault injection simulation selects critical faults from the dynamic slices, injects them at the specified time and evaluates the fault propagation.

\section{Experimental Results} \label{ExperimentalResults}

In order to verify the accuracy of proposed fault injection method, we firstly integrate our methodology into Cadence flow, then we execute our application on different designs that are available in~\cite{Clarke2002ProgramSF} and~\cite{opencores}. Table~\ref{bigtable} shows the details for both static slice which is state-of-the-art approach and dynamic slice optimization. For the smaller \emph{chopper} example, total CPU time of overall regression is reduced to 1.2s when compared to static slice optimizations. For the more complex \emph{simple\_spi} design, two-dimensional memory is selected as a fault target. As a result, we reduce the fault list to the critical faults and achieve 11.2 times shorter CPU time in dynamic slice optimization.

\section{Conclusions}

This paper proposes a methodology to optimize fault injection campaigns by pruning the fault list to the critical faults identified using a dynamic HDL slicing technique that provides for fault list collapsing. In this way, we narrow down the fault space and reduce execution time of fault injection simulation campaigns. Experimental results show that we achieve significant speed-up of the fault injection simulation when comparing to the state-of-the-art flows.

\bibliographystyle{IEEEtran}
\bibliography{ref}

\begin{thebibliography}{10}
\providecommand{\url}[1]{#1}
\csname url@samestyle\endcsname
\providecommand{\newblock}{\relax}
\providecommand{\bibinfo}[2]{#2}
\providecommand{\BIBentrySTDinterwordspacing}{\spaceskip=0pt\relax}
\providecommand{\BIBentryALTinterwordstretchfactor}{4}
\providecommand{\BIBentryALTinterwordspacing}{\spaceskip=\fontdimen2\font plus
\BIBentryALTinterwordstretchfactor\fontdimen3\font minus
  \fontdimen4\font\relax}
\providecommand{\BIBforeignlanguage}[2]{{%
\expandafter\ifx\csname l@#1\endcsname\relax
\typeout{** WARNING: IEEEtran.bst: No hyphenation pattern has been}%
\typeout{** loaded for the language `#1'. Using the pattern for}%
\typeout{** the default language instead.}%
\else
\language=\csname l@#1\endcsname
\fi
#2}}
\providecommand{\BIBdecl}{\relax}
\BIBdecl

\bibitem{ISO}
I.~S. Organization, ``Iso 26262 - road vehicles - functional safety,''
  \emph{International Organization for Standardization}, 2011.

\bibitem{7684076}
X.~Iturbe, B.~Venu, and E.~Ozer, ``Soft error vulnerability assessment of the
  real-time safety-related arm cortex-r5 cpu,'' in \emph{2016 IEEE
  International Symposium on Defect and Fault Tolerance in VLSI and
  Nanotechnology Systems (DFT)}, Sept 2016, pp. 91--96.

\bibitem{8347235}
R.~Travessini, P.~R.~C. Villa, F.~L. Vargas, and E.~A. Bezerra, ``Processor
  core profiling for seu effect analysis,'' in \emph{2018 IEEE 19th
  Latin-American Test Symposium (LATS)}, March 2018, pp. 1--6.

\bibitem{newJ}
J.~Raik, U.~Repinski, M.~Jenihhin, and A.~Chepurov, ``High-level decision
  diagram simulation for diagnosis and soft-error analysis,'' \emph{Design and
  Test Technology for Dependable Systems-on-Chip}, pp. 294--309, 2011.

\bibitem{6549113}
M.~Jenihhin, A.~Tšepurov, V.~Tihhomirov, J.~Raik, H.~Hantson, R.~Ubar,
  G.~Bartsch, J.~H.~M. Escobar, and H.~Wuttke, ``Automated design error
  localization in rtl designs,'' \emph{IEEE Design Test}, vol.~31, no.~1, pp.
  83--92, Feb 2014.

\bibitem{6233020}
U.~Repinski, H.~Hantson, M.~Jenihhin, J.~Raik, R.~Ubar, G.~D. Guglielmo,
  G.~Pravadelli, and F.~Fummi, ``Combining dynamic slicing and mutation
  operators for esl correction,'' in \emph{2012 17th IEEE European Test
  Symposium (ETS)}, May 2012, pp. 1--6.

\bibitem{Iwaihara96programslicing}
M.~Iwaihara, M.~Nomura, S.~Ichinose, and H.~Yasuura, ``Program slicing on vhdl
  descriptions and its applications,'' 1996.

\bibitem{Korel:1988:DPS:56378.56386}
\BIBentryALTinterwordspacing
B.~Korel and J.~Laski, ``Dynamic program slicing,'' \emph{Inf. Process. Lett.},
  vol.~29, no.~3, pp. 155--163, Oct. 1988. [Online]. Available:
  \url{http://dx.doi.org/10.1016/0020-0190(88)90054-3}
\BIBentrySTDinterwordspacing

\bibitem{Clarke2002ProgramSF}
E.~M. Clarke, M.~Fujita, S.~P. Rajan, T.~W. Reps, S.~Shankar, and
  T.~Teitelbaum, ``Program slicing for vhdl,'' \emph{International Journal on
  Software Tools for Technology Transfer}, vol.~4, pp. 125--137, 2002.

\bibitem{opencores}
\BIBentryALTinterwordspacing
(2018) Opencores. [Online]. Available: \url{http://www.opencores.org}
\BIBentrySTDinterwordspacing

\end{thebibliography}

\end{document}